\documentclass[floatfix,preprintnumbers,superscriptaddress,twocolumn]{revtex4}
\usepackage{amsmath,amssymb,color,epsfig,graphics,graphicx,lipsum,mathrsfs}

\begin{document}

\preprint{RESCEU-16/23}

\title{Quartic Gradient Flow}

\author{Muzi Hong}
\email{hong@resceu.s.u-tokyo.ac.jp}
\affiliation{Research Center for the Early Universe (RESCEU), Graduate School of Science, The University of Tokyo, Tokyo 113-0033, Japan}

\author{Ryusuke Jinno}
\email{ryusuke.jinno@resceu.s.u-tokyo.ac.jp}
\affiliation{Research Center for the Early Universe (RESCEU), Graduate School of Science, The University of Tokyo, Tokyo 113-0033, Japan}

\date{\today}

\begin{abstract}
Saddle-point configurations, such as the Euclidean bounce and sphalerons, are known to be difficult to find numerically.
In this Letter we study a new method, Quartic Gradient Flow, to search for such configurations.
The central idea is to introduce a gradient-flow-like equation in such a way that all the fluctuations around the saddle-point have eigenvalues that are square of the eigenvalues of the original quadratic operator.
We illustrate how the method works for the Euclidean bounce and sphalerons.
\end{abstract}

\maketitle

\vskip 0.5cm

\noindent
{\it Introduction.}
Saddle-point configurations often appear in physics.
One of the best known examples is the bounce solution that appears in the calculation of the tunneling rate of false vacuum decay~\cite{Coleman:1977py,Callan:1977pt}.
Another example is the sphaleron solution~\cite{Manton:1983nd,Klinkhamer:1984di} well-known for its important applications in baryogenesis ({\it e.g.}, Refs.~\cite{Arnold:1987mh,Kuzmin:1985mm,Fukugita:1986hr}).

A saddle point has negative mode(s), whereas a minimum has only positive modes.
This property of the former can cause problems when one tries to find it numerically.
For example, finding multi-field bounces numerically is a non-trivial problem, and a number of approaches have been proposed~\cite{Claudson:1983et,Kusenko:1995jv,Kusenko:1996jn,Dasgupta:1996qu,Moreno:1998bq,John:1998ip,Cline:1998rc,Cline:1999wi,Athron:2019nbd,Konstandin:2006nd,Park:2010rh,Wainwright:2011kj,Akula:2016gpl,Masoumi:2016wot,Espinosa:2018hue,Espinosa:2018szu,Jinno:2018dek,Piscopo:2019txs,Guada:2018jek,Chigusa:2019wxb,Sato:2019axv,Sato:2019wpo,Guada:2020xnz}.
In this context, the gradient flow method, which has been used in various fields of physics to solve differential equations, cannot be used at least naively due to these negative modes.
However, a recent proposal~\cite{Chigusa:2019wxb} showed that the effect of the negative mode(s) can be eliminated by adding an auxiliary term to the flow equation, and thus gradient flow can be used for the search of the bounce (see also Ref.~\cite{Sato:2019axv,Sato:2019wpo}).
The same technique was applied to sphalerons in Ref.~\cite{Hamada:2020rnp}.

In this Letter, we explore yet another approach for the search of saddle points using gradient flow.
We construct a flow equation in such a way that, instead of the original eigenvalues of the fluctuations, their squares appear.
As a result, all modes become positive (up to zero modes) and the flow converges to the desired solution.
This possibility is also mentioned in the footnote 4 of Ref.~\cite{Chigusa:2019wxb}.

The structure of the Letter is as follows.
We introduce the idea in Section 2, and show the numerical results for the calculation of the bounce solutions and sphaleron solutions in Section 3.
We discuss future applications and conclude in Section 4.

\vskip 0.5cm

\noindent
{\it Central idea.}
The central idea of this Letter is to introduce a gradient-flow-like equation in such a way that all the fluctuations around the saddle-point solution have positive eigenvalues alone.
We illustrate with a single real scalar field, but extension to multi-field is straightforward.
We introduce Quartic Gradient Flow as
\begin{align}
\partial_t \phi
+
\hat{\cal M} [\phi] \left[ \frac{\delta S}{\delta \phi} \right]
&=
0.
\label{eq:QGF}
\end{align}
Here $\hat{\cal M} [\phi] = \delta^2 S / \delta \phi \delta \phi = - \Delta + \cdots$, and $\phi = \bar{\phi} + \delta \phi$ denotes the saddle-point solution and a fluctuation around it.

To see how it works, let us expand the equation around the saddle point.
The action around the saddle point can be expanded to second order as
\begin{align}
S
&=
\int d^dx~
\left(
{\cal L}[\bar{\phi}]
+
\frac{1}{2} \delta \phi \, \hat{\cal M} [\bar{\phi}] \, \delta \phi
+
{\cal O} (\delta \phi^3)
\right),
\end{align}
and thus the flow reduces to
\begin{align}
\partial_t \phi
+
\hat{\cal M}^2 [\bar{\phi}] \delta \phi
&\simeq
0.
\end{align}
We may expand the fluctuation in terms of the eigenfunctions of $\hat{\cal M} [\bar{\phi}]$ as
\begin{align}
\delta \phi
&=
\sum_n c_n \delta \phi_n
\end{align}
where $c_1 \leq c_2 \leq c_3 \leq \cdots$ is assumed without losing generality, and the $n$-th eigenfunction and the $n$-th eigenvalue are denoted as $\delta \phi_n$ and $\lambda_n$, respectively.
Often there exist negative mode(s) $\lambda_1 < 0$.
The difficulty of naive gradient flow approaches lies in the negativity of this eigenvalue. 
However, because of the squared operator $\hat{\cal M}^2 [\bar{\phi}]$, Quartic Gradient Flow behaves as
\begin{align}
\partial_t \vec{c}
+
\Lambda^2 \vec{c}
&\simeq
0,
\qquad
\Lambda^2
=
{\rm diag} (\lambda_1^2, \lambda_2^2, \lambda_3^2, \cdots),
\label{eq:c}
\end{align}
for the coefficients $\vec{c} = (c_1, c_2, c_3, \cdots)$.
Since the square of a negative value is positive, the flow is expected to converge to the saddle-point solution up to zero-mode directions~\footnote{
Note in passing that the Quartic Gradient Flow method can be interpreted as gradient flow to the squared action
\begin{equation}
S'=\int d^d x \left(\frac{\delta S}{\delta \phi}\right)^2.
\end{equation}
\noindent
This action has been proposed in Ref.~\cite{Moreno:1998bq,John:1998ip} to find bounce solutions.
}~\footnote{One might be concerned about the effect of the zero mode(s). Zero modes represent the ``directions" along which the system is invariant, and thus $c_0$'s can take any real value. According to Eq.~(\ref{eq:c}), once $c_0$'s are fixed by the initial conditions, they do not change as the fictitious time $t$ evolves, and hence they do not spoil our method.}.

In the following we illustrate how Quartic Gradient Flow works, taking Euclidean bounce and sphalerons as examples.
We use forward Euler method to solve the partial differential equations.
See the Appendix~\ref{app:numerical} for numerical details.

\vskip 0.5cm

\begin{figure}
\centering 
\includegraphics[width=0.54\columnwidth]{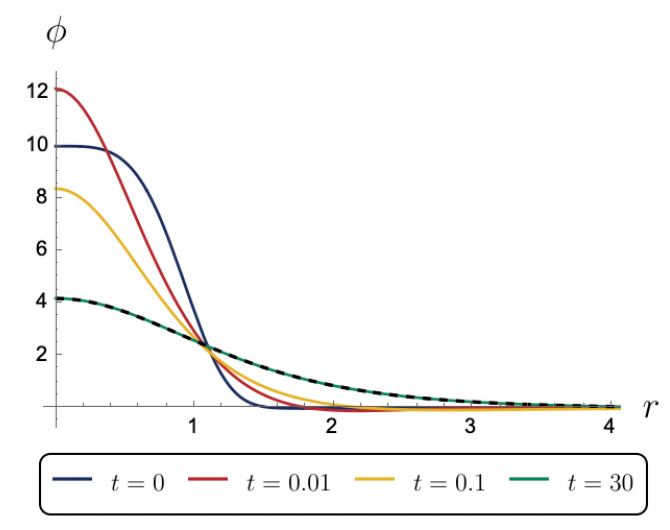}
\includegraphics[width=0.44\columnwidth]{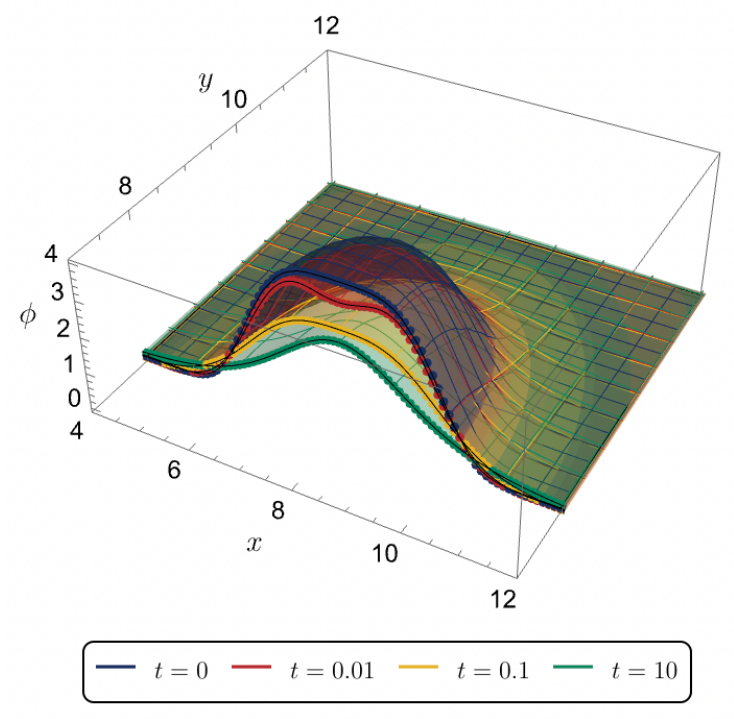}
\caption{
(Left) Behavior of Quartic Gradient Flow in Example 1 (single-field bounce) for $d = 3$ with a spherical symmetric setup.
The black dashed line is calculated with the overshoot-undershoot method.
(Right) Behavior of Quartic Gradient Flow without assuming spherical symmetry.
}
\label{fig:Example1}
\end{figure}

\noindent
{\it Example 1: single-field Euclidean bounce.}
We first consider the Euclidean bounce of a single real scalar field in $d$-dimensions.
In this case the functional $S$ is the Euclidean action $S_E$.
The bounce configuration is known to be $O(d)$ symmetric, and thus we may write the action as
\begin{align}
S_E
&=
\int d^d x~
\left[
\frac{1}{2} (\partial \phi)^2 + V (\phi)
\right]
\nonumber \\
&=
S_{d - 1}
\int r^{d - 1} dr~
\left[
\frac{1}{2} (\partial_r \phi)^2 + V (\phi)
\right].
\end{align}
Here the metric is ${\rm diag} (+, +, \cdots)$ in the first line, and $S_{d - 1}$ is the surface area of a $(d - 1)$-dimensional unit sphere.
The quadratic operator and the Euler-Lagrange equation are
\begin{align}
\hat{{\cal M}} [\phi]
&=
- \partial_r^2 - \frac{d - 1}{r} \partial_r + V'' (\phi),
\end{align}
and
\begin{align}
\frac{\delta S_E}{\delta \phi}
&=
- \partial_r^2 \phi - \frac{d - 1}{r} \partial_r \phi + V' (\phi)
=
0,
\end{align}
respectively, and thus Quartic Gradient Flow takes the form
\begin{align}
\partial_t \phi
+
\left[ \partial_r^2 + \frac{d - 1}{r} \partial_r - V''\right] \left[ \partial_r^2 \phi + \frac{d - 1}{r} \partial_r \phi - V' \right]
&=
0.
\end{align}
The left panel of Fig.~\ref{fig:Example1} shows how the initial configuration (blue) converges to the bounce for an example potential $V (\phi) = m^2 \phi^2 / 2 - a \phi^3 / 3$ with $m = a$ (taken to be unity without loss of generality).
For comparison, the black dashed line is the bounce solution calculated with the overshoot-undershoot method.
The value $\phi(0) \simeq 4.19$ coincides in both methods.

We may also remove the assumption of $O (d)$ symmetry.
In this case, Quartic Gradient Flow becomes
\begin{align}
\partial_t \phi
+
\left[ \Delta - V'' (\phi) \right] \left[ \Delta \phi - V' (\phi) \right]
&=
0.
\end{align}
In the right panel of Fig.~\ref{fig:Example1} we use the same potential as before and show how the configuration converges to the saddle point.
The value $\phi (x = L/2, y = L / 2) \simeq 2.39$ with $L=12$ coincides with the value calculated with the overshoot-undershoot method assuming spherical symmetry in $d = 2$ dimensions.

\vskip 0.5cm

\begin{figure}
\centering 
\includegraphics[width=\columnwidth]{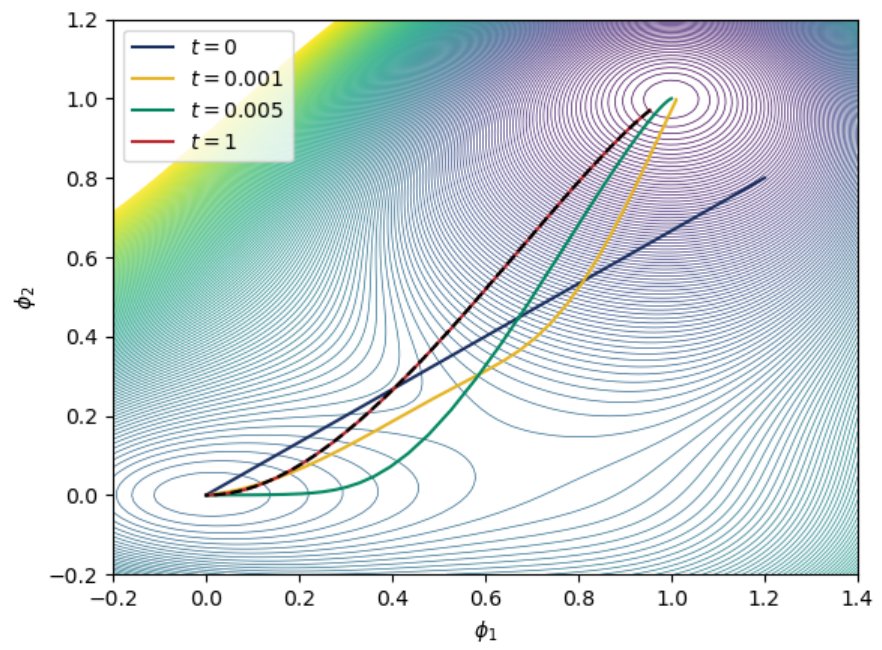}
\includegraphics[width=0.48\columnwidth]{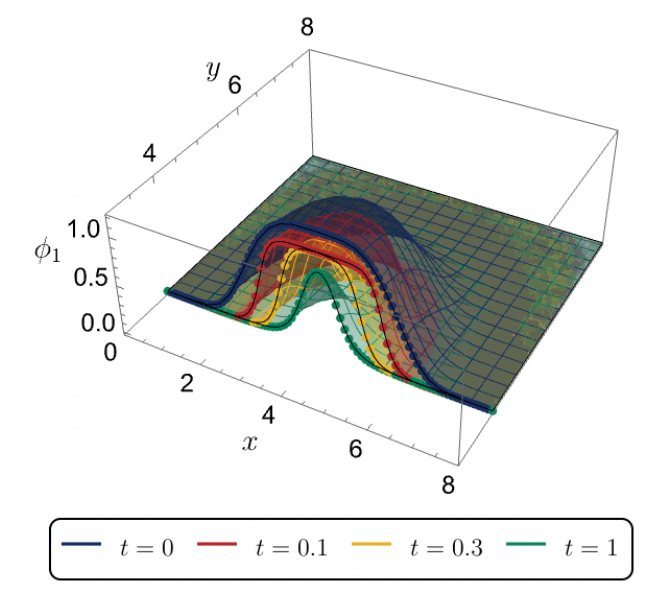}
\includegraphics[width=0.48\columnwidth]{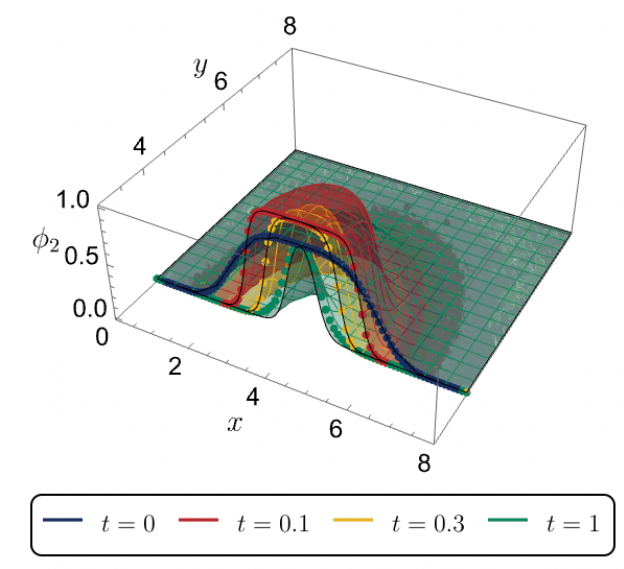}
\caption{
(Top) Behavior of Quartic Gradient Flow in Example 2 (multi-field bounce) for $d =3$ with a spherically symmetric setup.
The black dotted line is the result of \texttt{CosmoTransitions}.
(Bottom) Behavior of Quartic Gradient Flow for $d = 2$, without assuming spherical symmetry.
The left and right panels are $\phi_1$ and $\phi_2$, respectively.
}
\label{fig:Example2}
\end{figure}

\noindent
{\it Example 2: multi-field Euclidean bounce.}
We next consider a multi-field case with two real scalar fields.
Again the functional $S$ is identified with $S_E$, which we assume to be
\begin{align}
S_E
&=
\int d^d x~
\left[
\frac{1}{2} (\partial \phi_1)^2 + \frac{1}{2} (\partial \phi_2)^2 + V (\phi_1, \phi_2)
\right].
\end{align}
Quartic Gradient Flow becomes
\begin{align}
\partial_t \vec{\phi}
+
\hat{\cal M} [\vec{\phi}] \left[ \frac{\delta S_E}{\delta \vec{\phi}} \right]
&=
0,
\label{eq:QGF_multi}
\end{align}
with $\vec{\phi} = (\phi_1, \phi_2)^T$ and
\begin{align}
\hat{{\cal M}} [\vec{\phi}]
&=
\left(
\begin{array}{cc}
- \partial_r^2 - \frac{d - 1}{r} \partial_r + \partial_{\phi_1}^2 V
&
\partial_{\phi_1} \partial_{\phi_2} V
\\
\partial_{\phi_1} \partial_{\phi_2} V
&
- \partial_r^2 - \frac{d - 1}{r} \partial_r + \partial_{\phi_2}^2 V
\end{array}
\right),
\\
\frac{\delta S_E}{\delta \vec{\phi}}
&=
\left(
\begin{array}{c}
\displaystyle
- \partial_r^2 \phi_1 - \frac{d - 1}{r} \partial_r \phi_1 + \partial_{\phi_1} V
\\[0.2cm]
\displaystyle
- \partial_r^2 \phi_2 - \frac{d - 1}{r} \partial_r \phi_2 + \partial_{\phi_2} V
\end{array}
\right).
\end{align}
For the potential, we use an example in \texttt{CosmoTransitions}~\cite{Wainwright:2011kj}
\begin{align}
V (\phi_1, \phi_2)
&=
(\phi_1^2 + 5 \phi_2^2) [5 (\phi_1 - 1)^2 + (\phi_2 - 1)^2]
\nonumber \\
&\quad
+ k \left( \frac{1}{4} \phi_2^4 - \frac{1}{3} \phi_2^3 \right),
\end{align}
with $k = 80$.
The top panel of Fig.~\ref{fig:Example2} shows how Quartic Gradient Flow converges to the bounce solution.
The values $\phi_1 (0) \simeq 0.95$ and $\phi_2 (0) \simeq 0.97$ coincide with the values calculated with \texttt{CosmoTransitions}~\footnote{
We change the parameter \texttt{fRatioConv} in \texttt{CosmoTransitions} to 0.001 to improve accuracy as in Ref.~\cite{Chigusa:2019wxb}.
}.

Similarly as Example 1, we may also start without assuming $O (d)$ symmetry.
In this case, we can simply use
\begin{align}
\hat{{\cal M}} [\phi]
&=
\left(
\begin{array}{cc}
- \Delta + \partial_{\phi_1}^2 V
&
\partial_{\phi_1} \partial_{\phi_2} V
\\
\partial_{\phi_1} \partial_{\phi_2} V
&
- \Delta + \partial_{\phi_2}^2 V
\end{array}
\right),
\end{align}
and
\begin{align}
\frac{\delta S_E}{\delta \vec{\phi}}
&=
\left(
\begin{array}{c}
- \Delta \phi_1 + \partial_{\phi_1} V
\\
- \Delta \phi_2 + \partial_{\phi_2} V
\end{array}
\right),
\end{align}
in Eq.~(\ref{eq:QGF_multi}).
In the bottom panels of Fig.~\ref{fig:Example2} we show how the configuration converges to the bounce solution.
Note that the resulting configuration corresponds to the spherical bounce with $d = 2$.
The values $\phi_1 (x = L/2, y = L / 2) \simeq 0.80$ and $\phi_2 (x = L/2, y = L / 2) \simeq 0.79$ with $L=8$ coincide with those calculated with \texttt{CosmoTransitions} setting $d = 2$.

\vskip 0.5cm

\noindent
{\it Example 3: sphalerons.}
We also consider SU(2) sphalerons.
Sphalerons are stationary unstable solution of the gauge and Higgs fields.
The starting action is
\begin{align}
S
&=
\frac{1}{g^2} \int d^dx~
\left[
- \frac{1}{2} {\rm tr} \, F_{\mu \nu} F^{\mu \nu} + |D_\mu \Phi|^2 + V (\Phi)
\right],
\end{align}
and after imposing spherical symmetry and fixing the gauge, the energy becomes (see Appendix~\ref{app:sphaleron action} for derivation)
\begin{align}
E
=
\frac{4 \pi}{g^2}
\int_0^\infty dr
&\left[
|\partial_r \chi|^2
+
r^2 |\partial_r \phi|^2
+
\frac{1}{2 r^2} (|\chi|^2 -1)^2
\right.
\nonumber\\
&\left.
+
\frac{1}{2} (|\chi|^2 + 1) |\phi|^2
-
\textrm{Re}(\chi^* \phi^2)
\right.
\nonumber\\
&\left.
+
\frac{\lambda}{g^2} r^2 \left(|\phi|^2 - \frac{1}{2}g^2 v^2 \right)^2
\right],
\label{eq:sphaleron_energy}
\end{align}
where $\chi$ and $\phi$ are complex fields.

As far as ordinary sphalerons are considered, only the real parts of $\chi$ and $\phi$ take nonzero values~\cite{Klinkhamer:1984di,Yaffe:1989ms}.
Thus, for the purpose of finding the sphaleron solution, we may restrict them to be real and in that case the solution becomes a minimum of the energy~(\ref{eq:sphaleron_energy}), not a saddle point.
However, since our purpose of studying sphalerons here is to illustrate how Quartic Gradient Flow works in the presence of negative modes, we include the imaginary parts as well in the following analysis.
In this case the sphaleron becomes a saddle point of Eq.~(\ref{eq:sphaleron_energy}).
We also note that the full expression (\ref{eq:sphaleron_energy}) is necessary in finding the most relevant saddle-point solutions for larger $\lambda / g^2$ known as bisphalerons~\cite{Yaffe:1989ms}, though we restrict ourselves to ordinary sphalerons here.

We further redefine $\eta \equiv r \phi$ so that
\begin{align}
E'
&\equiv
\frac{g^2}{8 \pi} E
\nonumber \\
&=
\frac{1}{2}
\int_0^\infty dr
\left[
(\partial_r \chi_r)^2
+
(\partial_r \chi_i)^2
+
(\partial_r \eta_r)^2
+
(\partial_r \eta_i)^2
\right.
\nonumber\\
&\qquad \qquad \qquad
+
2 U (\chi, \eta; r) 
\left.
\right]
+ \textrm{const.},
\end{align}
with $\chi = \chi_r + i \chi_i$, $\eta = \eta_r + i \eta_i$, where the constant term comes from integration by parts, and
\begin{align}
U (\chi, \eta; r)
=
&\frac{1}{4 r^2} (|\chi|^2 -1)^2 + \frac{1}{4 r^2} (|\chi|^2 + 1) |\eta|^2
\nonumber\\
&- \frac{1}{2 r^2} \textrm{Re}(\chi^* \eta^2) + \frac{\lambda}{2 g^2 r^2} \left(|\eta|^2 - \frac{1}{2} g^2 v^2 r^2 \right)^2.
\end{align}
Then we may use Quartic Gradient Flow for $d = 1$, which takes
\begin{align}
\partial_t \vec{\phi}
+
\hat{\cal M} [\phi] \left[ \frac{\delta E'}{\delta \vec{\phi}} \right]
&=
0,
\label{eq:QGF_sphaleron}
\end{align}
with $\vec{\phi} = (\chi_r, \chi_i, \eta_r, \eta_i)^T$ and
\begin{widetext}
\begin{align}
\hat{{\cal M}} [\phi]
&= - \left(
\begin{array}{cccc}
\partial_r^2 - \partial_{\chi_r} U
&
- \partial_{\chi_r} \partial_{\chi_i} U
&
- \partial_{\chi_r} \partial_{\eta_r} U
&
- \partial_{\chi_r} \partial_{\eta_i} U
\\
- \partial_{\chi_i} \partial_{\chi_r} U
&
\partial_r^2 - \partial_{\chi_i} U
&
- \partial_{\chi_i} \partial_{\eta_r} U
&
- \partial_{\chi_i} \partial_{\eta_i} U
\\
- \partial_{\eta_r} \partial_{\chi_r} U
&
- \partial_{\eta_r} \partial_{\chi_i} U
&
\partial_r^2 - \partial_{\eta_r} U
&
- \partial_{\eta_r} \partial_{\eta_i} U
\\
- \partial_{\eta_i} \partial_{\chi_r} U
&
- \partial_{\eta_i} \partial_{\chi_i} U
&
- \partial_{\eta_i} \partial_{\eta_r} U
&
\partial_r^2 - \partial_{\eta_i} U
\end{array}
\right),
\qquad
\frac{\delta E'}{\delta \vec{\phi}}
=
- \left(
\begin{array}{c}
\partial_r^2 \chi_r - \partial_{\chi_r} U
\\
\partial_r^2 \chi_i - \partial_{\chi_i} U
\\
\partial_r^2 \eta_r - \partial_{\eta_r} U
\\
\partial_r^2 \eta_i - \partial_{\eta_i} U
\end{array}
\right).
\end{align}
\end{widetext}

\begin{figure}
\centering 
\includegraphics[width=0.48\columnwidth]{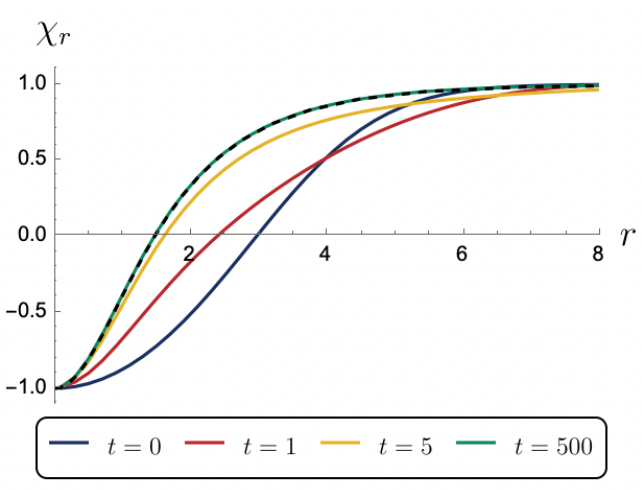}
\includegraphics[width=0.48\columnwidth]{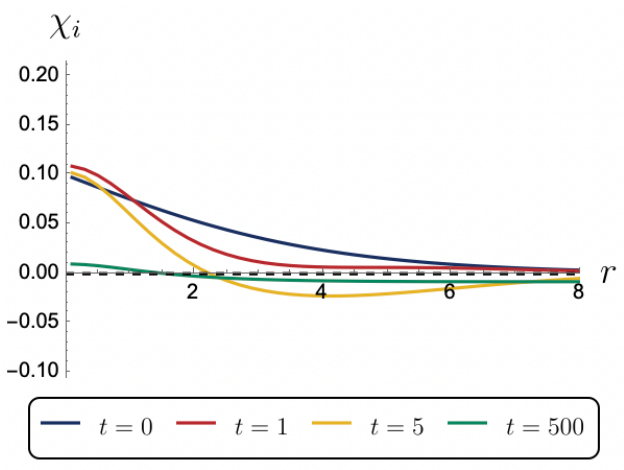}
\includegraphics[width=0.48\columnwidth]{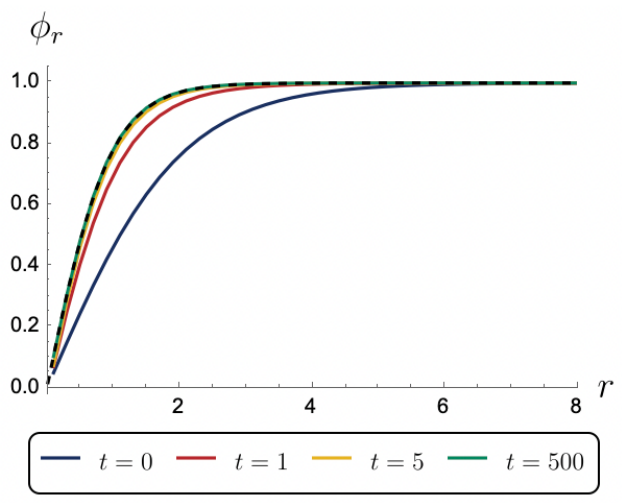}
\includegraphics[width=0.48\columnwidth]{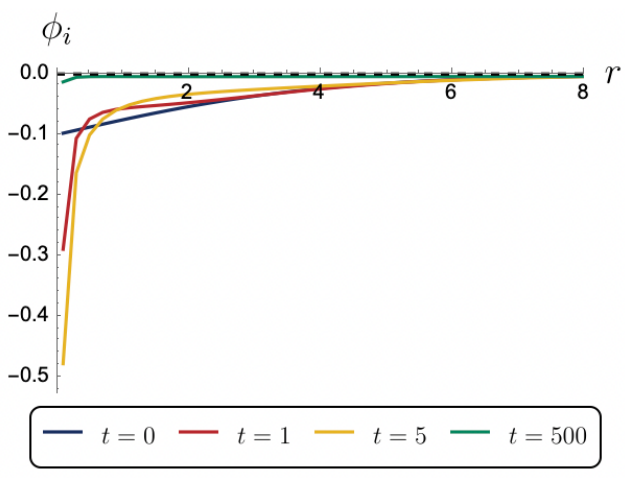}
\caption{
Behavior of Quartic Gradient Flow in Example 3 (sphaleron).
We take $\lambda / g^2 = 1$.
The left top, right top, left bottom, and right bottom panels are $\textrm{Re}\chi$, $\textrm{Im}\chi$, $\textrm{Re}\phi$ and $\textrm{Im}\phi$, respectively.
The black dashed line represents the results obtained with the numerical method in Ref.~\cite{Yaffe:1989ms}.
}
\label{fig:Example3}
\end{figure}

Fig.~\ref{fig:Example3} shows the behavior of the $\chi$ and $\phi$ fields in unit of $g v / \sqrt{2}$.
The colored lines are the time evolution of the fields, while the black dashed lines are the solutions obtained with the numerical method in Ref.~\cite{Yaffe:1989ms}.
We see that the fields converge to the desired sphaleron solution.
For the imaginary components, they should eventually converge to zero but there still remain errors of $O(0.01)$ at $t = 500$, possibly because of the finite time and number of grids.
In Table~\ref{table:ENCS} we also calculate the sphaleron energy for the converged solutions.

\begin{table}
\centering
\begin{tabular}{|c||c|c|c|c|} \hline
$\lambda / g^2$ & 0.1 & 0.5 & 1 & 5 \\ \hline
$E/(m_W/ \alpha_W)$ & 3.60 & 3.95 & 4.13 & 4.56 \\ \hline 
\end{tabular}
\caption{\small
Sphaleron energy $E$ calculated with Quartic Gradient Flow, evaluated at $t = 500$.
The results match those obtained with the numerical method of Ref.~\cite{Yaffe:1989ms}.
}
\label{table:ENCS}
\end{table}

\vskip 0.5cm

\noindent
{\it Discussion and Conclusions.}
In this Letter we study a novel method to calculate saddle-point configurations.
The central idea is to introduce a gradient-flow-like equation in such a way that all the fluctuations around the saddle point have positive eigenvalues in the flow equation, Eq.~(\ref{eq:QGF}).
This is made possible by the non-negativity of the eigenvalues of the operator $\hat{\cal M}^2 [\bar{\phi}]$ around the saddle point, with $\hat{\cal M} [\bar{\phi}]$ being the quadratic operator in the action.
The algorithm of Quartic Gradient Flow (\ref{eq:QGF}) is simple but powerful: it is applicable to general saddle-point searches without setting up different algorithms for different problems nor finding tailor-made ansatz.
We illustrate it with the bounce and sphaleron configurations. 

However, we also find several areas for improvement.
One of the major ones is the boundary condition.
As detailed in the Appendix~\ref{app:numerical}, for $O(d)$ bounce solutions we fix the value at the origin by imposing a boundary condition consistent with the original bounce equation (see the text below Eqs.~(\ref{eq:d1phi})--(\ref{eq:d4phi})).
Without it the configuration would pass through the bounce and reach the trivial solution.
Part of the reason for this behavior might be the forward Euler method we use in this Letter.
It may cause divergence near the origin in the spherical coordinate when the potential contains terms involving divergent factors such as $\sim 1/r$.
We confirmed that this behavior does not occur if one uses backward Euler method instead.
Also, as shown in Fig.~\ref{fig:Example3} for the sphaleron case, the forward Euler method cannot fix the value at the origin.
Although one can always use boundary conditions proposed in Refs.~\cite{Manton:1983nd,Klinkhamer:1984di,Yaffe:1989ms} to speculate the field value at the origin, we expect advanced numerical method can calculate the full configuration accurately without additional assumptions.

We present possible future directions from the viewpoint of both numerical methods and physical applications.
As mentioned above, explicit numerical methods such as the forward Euler method may bring divergence if the problem involves spherical coordinates.
In contrast, implicit methods such as the backward Euler and the Crank-Nicolson methods generally do not lead to such divergence.
Implementing the latter methods may allow for convergence of the configuration to the desired solution without some of the boundary conditions imposed in this study.
As for physical applications, one may think of various saddle point searches.
For example, one could try the method for thin-wall bounce, which we did not consider in this Letter.
Sphalerons also have many possible applications.
Sphaleron configurations with Chern-Simons number different from $1/2$, called bisphalerons, are known to appear when $m_H^2 / m_W^2$ is larger than the Standard Model 
value~\cite{Yaffe:1989ms}, and their properties might be investigated with Quartic Gradient Flow.
It would also be interesting to study the properties of sphalerons with a finite Weinberg angle, see {\it e.g.}, Refs.~\cite{James:1992re,Klinkhamer:1990fi,Kunz:1992uh}.
Moreover, other types of sphalerons exist in the Standard Model such as $S^*$ and $\hat{S}$~\cite{Klinkhamer:1993hb,Klinkhamer:2003hz}.
Exploring these configurations using Quartic Gradient Flow would be exciting possibilities, which we leave for future work.

\vskip 0.5cm

\noindent
{\it Acknowledgments.}
The authors are grateful to Yu Hamada, Masazumi Honda, Takeo Moroi, Yutaro Shoji and Hiromasa Watanabe for helpful comments.
R.~J. is supported by donuts sold in seven-eleven.
M.~H. is supported by FoPM, WINGS Program, the University of Tokyo and Grant-in-Aid for JSPS Fellows 23KJ0697.

\pagebreak
\widetext
\vskip 0.2cm
\begin{center}
\textbf{\large Supplemental Material: Quartic Gradient Flow}
\end{center}

\setcounter{equation}{0}
\setcounter{figure}{0}
\setcounter{table}{0}
\setcounter{page}{1}
\makeatletter
\renewcommand{\theequation}{S\arabic{equation}}
\renewcommand{\thefigure}{S\arabic{figure}}

\section{Numerical details}
\label{app:numerical}

In this Supplemental Material we summarize numerical details to reproduce the results in the Letter.

\vskip 0.5cm

\noindent
{\it Example 1: single-field Euclidean bounce.}
The first example is the single-field bounce.
For the case with spherical symmetry, we take $r$-grid as $r_i = (i / n) R$ ($i = 0, 1, \cdots, n$), and set a mirror at $r_0$ such that $\phi_{- 1} = \phi_1$ and $\phi_{- 2} = \phi_2$ with $\phi_i \equiv \phi (r_i)$.
We use
\begin{align}
\left[ \partial_r \phi \right]_i
&=
\frac{\phi_{i + 1} - \phi_{i - 1}}{2 \Delta r} + {\cal O} (\Delta r^2),
\label{eq:d1phi}
\\
\left[ \partial_r^2 \phi \right]_i
&=
\frac{ \phi_{i + 1} - 2 \phi_i + \phi_{i - 1}}{\Delta r^2} + {\cal O} (\Delta r^2),
\label{eq:d2phi}
\\
\left[ \partial_r^3 \phi \right]_i
&=
\frac{\phi_{i + 2} - 2 \phi_{i + 1} + 2 \phi_{i - 1} - \phi_{i - 2}}{2 \Delta r^3} + {\cal O} (\Delta r^2),
\label{eq:d3phi}
\\
\left[ \partial_r^4 \phi \right]_i
&=
\frac{\phi_{i + 2} - 4 \phi_{i + 1} + 6 \phi_i - 4 \phi_{i - 1} + \phi_{i - 2}}{\Delta r^4} + {\cal O} (\Delta r^2)
\label{eq:d4phi}
\end{align}
\noindent
as finite difference. We use forward Euler method to obtain the results shown above. 
Care must be taken for the boundary conditions at $r = 0$.
The mirror already fixes $[\partial_r \phi]_0 = [\partial_r^3 \phi]_0 = 0$.
Forward Euler method causes numerical divergence around $r = 0$ in this system unless one takes extremely small $\Delta t$, and hence we calculate $\phi_0$ using an additional boundary condition to avoid it.
Since the bounce solution behaves as $\phi = \phi_0 + (1 / 2 d) (\partial V / \partial \phi |_{\phi = \phi_0}) r^2 + \cdots$ around $r \sim 0$, we further impose $[\partial_r^2 \phi]_0 = (1 / d) (\partial V / \partial \phi )_0$ at each step to fix the value of $\phi$ at $r = 0$.
When calculating $\phi_0$ here, we use 
\begin{equation}
\left[ \partial_r^2 \phi \right]_i
=
\frac{- \phi_{i + 2} + 16 \phi_{i + 1} - 30 \phi_i + 16 \phi_{i - 1} - \phi_{i - 2}}{12 \Delta r^2} + {\cal O} (\Delta r^4)
\label{eq:d2higheraccuracy}
\end{equation}
\noindent
as finite difference. This means that $\phi_0$ is updated according to $\phi_0 = [32 \phi_1 - 2 \phi_2 - 12 \Delta r^2 (1 / d) (\partial V / \partial \phi)_0] / 30$.
We observe no significant change in the result for the case of single-field bounce if one uses~(\ref{eq:d2phi}) to calculate $\phi_0$.
However, as elaborated below, the result of multi-field bounce slightly changes, and thus we take~(\ref{eq:d2higheraccuracy}) to calculate $\phi_0$.
We also impose $\phi_{n} = 0$ at each time step.
For terms involving $r$-derivatives of the potential, we do not expand them using the chain rule but rather evaluate $r$-derivatives directly, {\it e.g.} $\partial_r V' (\phi) = \partial_r [(V' \circ \phi) (r)]$.
With this setup, the left panel of Fig.~\ref{fig:Example1} is obtained with $d = 3$, $n = 200$, $R = 8$, $\Delta r = 0.04$, and $\Delta t = 10^{- 7}$.
The initial condition is taken as $\phi_{\rm ini} (r) = 10 e^{- r^4}$.

\begin{figure}
\centering 
\includegraphics[width=0.3\columnwidth]{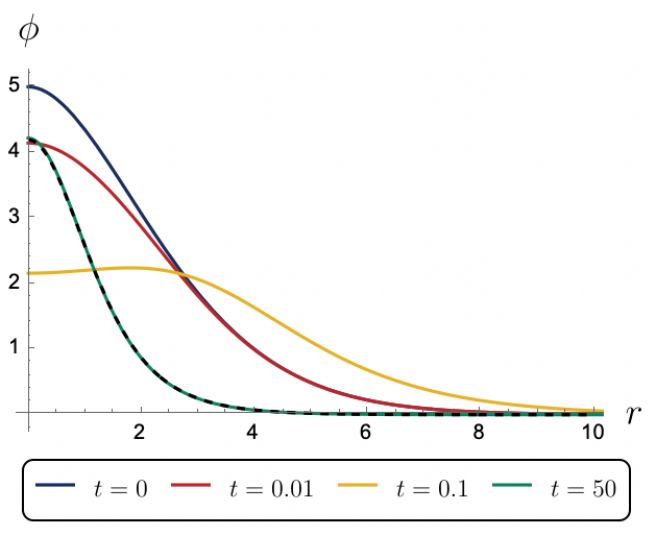}
\caption{
Behavior of Quartic Gradient Flow in Example 1 (single-field bounce) for $d = 3$ with a spherical symmetric setup using backward Euler method.
The black dashed line is calculated with the overshoot-undershoot method.}
\label{fig:appendix}
\end{figure}

We also calculate this case using the backward Euler method, which does not require fixing $\phi$ at $r = 0$ with the boundary condition.
We first rescale $r$ as in Ref.~\cite{Chigusa:2019wxb} into $x \in (0,1)$ with $x = \textrm{tanh}(r/R_0)$, where we set $R_0 = 6$.
We approximately set $x_i = 10^{-7} + (i/n)(1-2 \times 10^{-7}) (i = 0,1,\cdots,n)$.
We use the Gaussian elimination method to solve the linear algebraic equations at each time step.
When taking $d=3$, $n=100$ and $\Delta t = 10^{-5}$, one obtains $\phi_0 \simeq 4.22$ at $t = 50$. As shown in Fig.~\ref{fig:appendix}, 
the configuration first passes through the bounce solution but then comes back and converges to the bounce.
The initial condition is taken as $\phi_{\rm ini}(x) = 5 + 6 \times 10^{-6} x - 30 x^2 + 40 x^3 - 15 x^4$, so that the first derivative of $\phi_0 = \phi (x = 10^{-7})$ and $\phi_n = \phi (x = 1-10^{-7})$ is zero. We obtain $\phi_0 \simeq 4.20$ with $n=160$ and other setups unchanged.

For the case without spherical symmetry, we take $xy$-grid as $x_i = (i / n) L$ and $y_j = (j / n) L$ ($i, j = 0, 1, \cdots, n$).
We use
\begin{align}
\left[ (\partial_x^2 + \partial_y^2) \phi \right]_{i, j}
&=
\frac{\phi_{i + 1, j} - 2 \phi_{i, j} + \phi_{i - 1, j}}{\Delta d^2} + \frac{\phi_{i, j + 1} - 2 \phi_{i, j} + \phi_{i, j - 1}}{\Delta d^2} + {\cal O} (\Delta d^2),
\\
\left[ (\partial_x^2 + \partial_y^2)^2 \phi \right]_{i, j}
&=
\frac{\phi_{i + 2, j} - 4 \phi_{i + 1, j} + 6 \phi_{i, j} - 4 \phi_{i - 1, j} + \phi_{i - 2, j}}{\Delta d^4}
+
\frac{\phi_{i, j + 2} - 4 \phi_{i, j + 1} + 6 \phi_{i, j} - 4 \phi_{i, j - 1} + \phi_{i, j - 2}}{\Delta d^4}
\nonumber \\
&+
\frac{2 (\phi_{i + 1, j + 1} - 2 \phi_{i + 1, j} + \phi_{i + 1, j - 1}
- 2 \phi_{i, j + 1}
+ 4 \phi_{i, j}
- 2 \phi_{i, j - 1}
+ \phi_{i - 1, j + 1} - 2 \phi_{i - 1, j} + \phi_{i - 1, j - 1}
)}{\Delta d^4}
\nonumber \\
&+
{\cal O} (\Delta d^2),
\end{align}
as finite difference.
For the boundary condition, we impose $\phi_{i, 0} = \phi_{i, n} = \phi_{0, j} = \phi_{n, j} = 0$ at each time step.
With this setup, the right panel of Fig.~\ref{fig:Example1} is obtained with $n = 160$, $L = 16$, $\Delta d = 0.1$, and $\Delta t = 10^{- 6}$.
The initial condition is taken as $\phi_{\rm ini} (x, y) = 4 e^{- [(x - L / 2)^2 + (y - L / 2)^2]^2 / 16}$.

\vskip 0.5cm

\noindent
{\it Example 2: multi-field Euclidean bounce.}
For multi-field cases, the procedure is almost the same as single-field cases.
Note that, taking two-field cases for instance, cross terms $\partial_{\phi_1} \partial_{\phi_2} V \times [- \Delta \phi_2 + \partial_{\phi_2} V]$ and $\partial_{\phi_1} \partial_{\phi_2} V \times [- \Delta \phi_1 + \partial_{\phi_1} V]$ appear in the time evolution of $\phi_1$ and $\phi_2$, respectively.
With this setup, the top panel of Fig.~\ref{fig:Example2} is obtained with $d = 3$, $n = 160$, $R = 8$, $\Delta r = 0.05$, and $\Delta t = 10^{- 8}$.
The initial condition is taken as $\phi_{1, {\rm ini}} (r) = 1.2 e^{- r^4 / 16}$ and $\phi_{2, {\rm ini}} (r) = 0.8 e^{- r^4 / 16}$.
As mentioned above, if one calculates the field values at origin using~(\ref{eq:d2phi}) instead of Eq.~(\ref{eq:d2higheraccuracy}), one obtains $\phi_1(0) \simeq 0.96$ and $\phi_2(0) \simeq 0.97$ at $t = 5$ with $d = 3$, $n = 200$, $R = 8$, $\Delta r = 0.04$, and $\Delta t = 10^{- 7}$.
These values deviates by O(1\%) from the value obtained with \texttt{CosmoTransitions}.
Also, the bottom panels are obtained with $n = 80$, $L = 8$, $\Delta r = 0.1$, and $\Delta t = 10^{- 6}$.
The initial conditions are taken as $\phi_{1, {\rm ini}} (x, y) = 1.2 e^{- [(x - L / 2)^2 + (y - L / 2)^2]^2 / 16}$ and $\phi_{2, {\rm ini}} (x, y) = 0.8 e^{- [(x - L / 2)^2 + (y - L / 2)^2]^2 / 16}$.

\vskip 0.5cm

\noindent
{\it Example 3: sphaleron.}
For the case of sphalerons, we cannot naively use the same update procedure as Example 1, $\phi_0 = [32 \phi_1 - 2 \phi_2 - 12 \Delta r^2 (1 / d) (\partial U / \partial \phi)_0] / 30$ (with $d = 1$ and $\phi = \chi$ or $\eta$), as the potential term contains inverse powers of $r$.
To avoid substituting $r = 0$ to the potential, we take $r$-grid as $r_i = ((i - 1/2)/n) R \ (i = 1, \cdots, n)$, and set a mirror at $r = 0$ such that $\phi_{- 1} = \phi_1$ and $\phi_{- 2} = \phi_2$.
Note that we do not have $r_0$ in this setup and $r_1-r_{- 1} = R/n$.
We also impose $\eta_{n-1} = 2 \eta_{n-2} -\eta_{n-3}$ and $\eta_{n} = 2 \eta_{n-1} -\eta_{n-2}$ at each step, though this does not affect the final result very much.
With this setup, Fig.~\ref{fig:Example3} is obtained with $d = 1$, $n = 60$, $R = 12$, $\Delta r = 0.2$, and $\Delta t = 10^{- 5}$.
The initial condition is taken as $\chi_r (r) = - 1 + 2 \tanh (r^2 / 16)$, $\chi_i (r) = 0.1 - 0.1 \tanh (r / 4)$, $\eta_r (r) = r \tanh (r / 2)$, and $\eta_i (r) = r (-0.1 + 0.1 \tanh (r / 4))$.
When calculating energy given in Table~\ref{table:ENCS}, we use
\begin{equation}
\left[ \partial_r \phi \right]_i
=
\frac{- \phi_{i + 2} + 8 \phi_{i + 1} - 8 \phi_{i - 1} + \phi_{i - 2}}{12 \Delta r} + {\cal O} (\Delta r^4)
\end{equation}
as finite difference to calculate first derivatives.
We first calculate the value of $\phi(r)$ from $\phi = \eta / r$, and use~(\ref{eq:sphaleron_energy}) with imaginary fields set to zero to calculate energy.

\vskip 0.5cm

\section{The energy of sphalerons}
\label{app:sphaleron action}

We follow the discussion in Ref.~\cite{Yaffe:1989ms} to derive (\ref{eq:sphaleron_energy}). The action of SU(2)-Higgs theory is
\begin{equation}
S
= \frac{1}{g^2} \int d^4 x
\left[
- \frac{1}{2} \textrm{tr} F_{\mu \nu} F^{\mu \nu}
+
|D_\mu \Phi|^2
+ \frac{\lambda}{g^2} \left(\Phi^\dagger \Phi - \frac{1}{2} g^2 v^2 \right)^2
\right].
\label{eq:SU2Higgs_action}
\end{equation}
\noindent
The SU(2) gauge field is $A_\mu \equiv A^a_\mu (\tau^a/2i)$, where $\tau^a\ (a = 1,\ 2,\ 3)$ are the Pauli matrices, and $F_{\mu\nu} \equiv [D_\mu, D_\nu]$ is the SU(2) field strength.
$\Phi$ is an SU(2) doublet scalar field with covariant derivative $D_\mu \equiv (\partial_\mu + A_\mu)$.
We follow Yaffe's notation and use $\eta_{\mu \nu} \equiv \textrm{diag}(-,+,+,+)$ in Minkowski spacetime.
Here, we neglect U(1)$_Y$ field~\cite{Klinkhamer:1984di}.

When an O(3) rotation of spatial directions can be canceled by a combination of SU(2) gauge and SU(2) global transformations, the field configurations are called ``spherically symmetric."
An ansatz for the most general spherically symmetric configuration is given by Refs.~\cite{Witten:1976ck, Ratra:1987dp}
\begin{align}
A_0(x)
&=
[a_0(r,t) \vec{\tau} \cdot \hat{\vec{x}}] / 2i,
\\
A_j(x)
&=
\{[\alpha(r,t) - 1] e_j^1 / r
+
\beta(r,t) e_j^2 / r
+
a_1(r,t) e_j^3\} / 2i,
\\
\Phi(x)
&=
[\mu(r,t)
+
i \nu(r,t) \vec{\tau} \cdot \hat{\vec{x}}] \xi,
\end{align}
\noindent
where $\hat{\vec{x}} \equiv \vec{x} / r$, $r \equiv |\vec{x}|$, $\alpha(r,t), \beta(r,t), a_0(r,t), a_1(r,t), \mu(r,t), \nu(r,t)$ are arbitrary real functions and $\xi$ is an arbitrary two-component complex unit vector. $\{e^k_j\}$ are defined as
\begin{align}
e^1_i
&=
(\vec{\tau} \cdot \hat{\vec{x}} \tau_i
-
\hat{x}_i) / i
=
(\epsilon_{ijk} \hat{x}_k) \tau_j, \\
e^2_i
&=
\tau_i
-
\vec{\tau} \cdot \hat{\vec{x}} \hat{x}_i
=
(\delta_{ij}
-
\hat{x}_i \hat{x}_j) \tau_j, \\
e^3_i
&=
\vec{\tau} \cdot \hat{\vec{x}} \hat{x}_i
=
(\hat{x}_i \hat{x}_j) \tau_j.
\end{align}
\noindent
$(1+1)$-dimensional complex fields $\chi \equiv \alpha + i\beta$, $\phi \equiv \mu + i\nu$ and the $(1+1)$-dimensional field strength $f_{\mu \nu} = \partial_\mu a_\nu - \partial_\nu a_\mu$ with indices $\mu, \nu = 0, 1$ are defined for convenience. $(1+1)$-dimensional covariant derivatives are defined as
\begin{align}
D_\mu \chi 
&= 
(\partial_\mu - ia_\mu) \chi, 
\\
D_\mu \phi
&=
(\partial_\mu - ia_\mu / 2) \phi.
\end{align}
\noindent
This ansatz is consistent with a U(1) gauge transformation, which is a subgroup of the SU(2) gauge group, consisting of $\{\Omega = \textrm{exp}[i \omega(r,t) \vec{\tau} \cdot \hat{\vec{x}} / 2]\}$. Under these transformations,
\begin{align}
\chi
&\to
e^{i\omega} \chi, \\
\phi
&\to
e^{i\omega/2} \phi, \\
a_\mu
&\to
a_\mu + \partial_\mu \omega.
\end{align}
The action becomes
\begin{align}
S 
= 
&\frac{4\pi}{g^2} \int dt dr 
\left[ 
\frac{1}{4} r^2 f_{\mu\nu} f^{\mu\nu} 
+ 
|D\chi|^2 
+ 
r^2 |D\phi|^2 
- 
\textrm{Re} (\chi^* \phi^2) 
+ 
\frac{1}{2r^2} (|\chi|^2 - 1)^2 
\right.
\\
&\left.+ 
\frac{1}{2} (|\chi|^2 + 1) |\phi|^2 
+ 
\frac{\lambda}{g^2} r^2 \left(|\phi|^2 - \frac{1}{2} g^2 v^2 \right)^2 \right],
\end{align}
\noindent
using the ansatz given above.

One may first obtain strictly static configurations, which do not depend on time, with gauge fixing $a_1=0$.
After gauge transforming them one obtains arbitrary static spherically symmetric field configurations, which are static up to gauge transformations:
\begin{align}
\chi(r,t)
&=
e^{i\omega(r,t)} \bar{\chi}(r),
\\
\phi(r,t)
&=
e^{i\omega(r,t) /2 } \bar{\phi}(r),
\\
a_0(r,t)
&=
\partial_0 \omega(r,t) + \bar{a_0}(r),
\\
a_1(r,t)
&=
\partial_1 \omega(r,t).
\end{align}
\noindent
The energy of these static configurations is
\begin{align}
E
=
&\frac{4 \pi}{g^2} \int dr
\left[
\frac{1}{2} r^2 (\partial_r \bar{a}_0)^2
+
\bar{a}_0^2 \left( |\bar{\chi}|^2 + \frac{1}{4} r^2 |\bar{\phi}|^2 \right)
+
|\partial_r \bar{\chi}|^2
+
r^2 |\partial_r \bar{\phi}|^2
+
\frac{1}{2 r^2} (|\bar{\chi}|^2 - 1)^2
\right.
\\
&\left.
+
\frac{1}{2}(|\bar{\chi}|^2 + 1) |\bar{\phi}|^2
-
\textrm{Re}(\bar{\chi}^* \bar{\phi}^2)
+
\frac{\lambda}{g^2} r^2 \left(|\bar{\phi}|^2 - \frac{1}{2} g^2 v^2 \right)^2
\right].
\end{align}
\noindent
One of the field equations yields Gauss's law
\begin{align}
\left(
- \partial_r r^2 \partial_r
+
2 |\bar{\chi}|^2
+
\frac{1}{2} r^2 |\bar{\phi}|^2
\right)
\bar{a}_0
=
0.
\end{align}
\noindent
The operator before $\bar{a}_0$ is strictly positive, and thus $\bar{a}_0$ vanishes, and (\ref{eq:sphaleron_energy}) is obtained with $\bar{\chi}$, $\bar{\phi}$ replaced by $\chi$, $\phi$, respectively.

\bibliography{ref}

\end{document}